\documentstyle[prl,aps,epsf]{revtex}
\begin{document}
\draft
\twocolumn[\hsize\textwidth\columnwidth\hsize\csname @twocolumnfalse\endcsname
\title {Polaron effective mass from Monte Carlo simulations}
\author{P.\,E.\,Kornilovitch and E.\,R.\,Pike}
\address{Department of Physics, King's College London, Strand, 
London WC2R 2LS, UK}
\date{7 January 1997}
\maketitle
\begin{abstract}
A new Monte Carlo algorithm for calculating polaron 
effective mass is proposed. It is based on the
path-integral representation of a partial partition function 
with fixed total quasi-momentum. Phonon degrees of freedom
are integrated out analytically resulting in a single-electron 
system with retarded self-interaction and open boundary
conditions in imaginary time. The effective mass is
inversely proportional to the covariance of total energy
calculated on an electron trajectory
and the square distance between ends of the trajectory.
The method has no limitations on values of model
parameters and on the size and dimensionality of the
system although large statistics is required for 
stable numerical results. The method is tested 
on the one-dimensional Holstein model for which simulation
results are presented.   

\end{abstract}
\pacs{PACS numbers: 63.20.Kr, 71.38.+i}
\vskip2pc]
\narrowtext

The polaron problem is a classic problem of 
condensed-matter physics \cite{Appel,Alexandrov_one}. 
In recent years
it has attracted additional attention in relation to
high-temperature superconductivity \cite{Alexandrov_two}.
Currently active theoretical research is
being conducted by various methods on models with
strong electron-phonon phonon interaction as well as on
models with both electron-phonon and electron-electron
interaction such as Hubbard-Holstein and $t-J$-Holstein 
models. The Monte Carlo simulation method
is widely used in these studies \cite{Noack} 
because it provides unbiased information about 
thermodynamic and dynamic properties of the system even 
in regions of model parameters
hardly accessible by standard theoretical tools.

The polaron problem was among the first applications of
the Quantum Monte Carlo method after its introduction 
in condensed-matter physics in the
Seventies. In simulations of many-polaron systems   
either both the positions of the electrons and the 
ions are simulated \cite{Hirsch_one,Hirsch_two} or 
the electrons are integrated out analytically and
only the phonon subsystem is simulated \cite{White}.
For the single-polaron problem,
De Raedt and Lagendijk \cite{DeRaedt} developed an effective 
algorithm in which phonon degrees of freedom
are integrated out analytically by the Feynman
technique \cite{Feynman}. This results in a single-electron
problem with retarded self-interaction along
the imaginary-time direction. Since a system with
only one degree of freedom can be easily simulated 
by the Metropolis Monte Carlo algorithm 
\cite{Metropolis}  
this method provides accurate results
for thermodynamic polaron properties such as   
internal energy and specific heat and on other 
static quantities of interest, for instance the electron-phonon
correlation function. Unfortunately, the estimation
of dynamic properties with this method, 
such as the polaron spectrum,
is much more difficult, the reason being that this 
would require an analytic continuation of simulation
results from the imaginary-time to the real-time domain.
This procedure is mathematically ill-posed and is
thus very sensitive to 
the statistical noise in the input data.
In previous work \cite{Creffield} we have applied
the singular-value decomposition
technique to carry out such a continuation. However,
due to the narrowing effect \cite{Appel,Alexandrov_one} 
the polaron band is expected to be very narrow, 
especially in the strong-coupling regime, and the bandwidth 
smaller than the accuracy with which the polaron spectrum
could be obtained by analytic continuation.

In this paper we show that despite these difficulties
the polaron effective mass, which is defined as
an inverse second derivative of internal energy
with respect to quasi-momentum in the limit of zero
temperature, can be obtained directly from Monte Carlo
simulations without going to real time. This possibility
is based on a special kind of fluctuation-dissipation
relation and resembles Feynman's calculation 
of the Fr\"ohlich polaron effective mass \cite{Feynman}
and the method of calculation  
of the superfluid fraction in liquid helium \cite{Ceperley}. 

Our starting point is the fact that total quasi-mo\-men\-tum
${\bf P}$ is a conserved quantity in an electron-phonon
system. Since states with different ${\bf P}$ do not mix
during the evolution it is meaningful to calculate
a partial partition function $Z_{\bf P}$ which is a sum
over states with fixed ${\bf P}$
\begin{equation}
Z_{\bf P} = \sum_n \langle n | e^{- \beta H} | n \rangle
\, \delta_{{\bf P},{\bf P}_n} .
\label{one}
\end{equation}
Here $H$ is the hamiltonian, $\beta = 1/k_B T$, $n$ numbers
states of a complete basis in the Hilbert space 
and ${\bf P}_n$ is the total quasi-momentum
of state $|n\rangle$. The total partition function is
obviously $Z=\sum_{\bf P} Z_{\bf P}$.
The total Hilbert space
is a direct product of the one-electron $\{ {\bf k} \}$
and the phonon Hilbert spaces. The latter in turn can
be divided into subspaces $\{ n^{\bf Q}_{ph} \}$ with fixed total phonon
quasi-momentum ${\bf Q}$. We rewrite (\ref{one}) as follows
\begin{equation}
Z_{\bf P} = \sum_{\bf k} \sum_{\bf Q} \sum_{n^{\bf Q}_{ph} } 
\langle {\bf k}, n^{\bf Q}_{ph} | e^{- \beta H} |
n^{\bf Q}_{ph} , {\bf k} \rangle \, \delta_{{\bf P, \, k+Q}} . 
\label{two}
\end{equation}
We have used a momentum-space basis because 
the conservation of quasi-momentum is conveniently written 
in momentum space. However, matrix elements of the evolution 
operator ${\rm exp}(-\beta H)$ are easily calculated in
real space \cite{DeRaedt}. Therefore, our first goal is to 
transform Eq.(\ref{two}) to a real-space basis. We do
this in two steps. Firstly, introducing the  
Wannier basis for the electron $|{\bf r} \rangle =
N^{-1/2} \sum_{\bf k} e^{i {\bf k r}} |{\bf k} \rangle$
and exercising the delta-function one gets
\begin{displaymath}
Z_{\bf P} = \frac{1}{N} \sum_{\bf r,r'} \sum_{\bf Q}
e^{i{\bf (P-Q)(r'-r)}}
\sum_{n^{\bf Q}_{ph} }
\langle {\bf r'}, n^{\bf Q}_{ph} | e^{- \beta H} |
n^{\bf Q}_{ph} , {\bf r} \rangle 
\end{displaymath}
where $N$ is the total number of sites in the lattice.
Secondly, we introduce a new basis for the phonon states 
in terms of ion configurations 
$|\{ \xi_i \} \rangle $, $i=1, \ldots, N$.
Such a configuration
represents a state where $i$-th ion is displaced
from its equilibrium position by distance $\xi_i$.
(For simplicity we assume only one phonon degree
of freedom per lattice site).
These states are normalised as follows
$\langle \{ \xi_i \} | \{ \xi'_i \} \rangle =
\prod_{i=1}^N \delta(\xi_i - \xi'_i)$ and
resolve the identity: ${\bf I} = \int_{-\infty}^{\infty} \! D\xi
\, |\{ \xi_i \} \rangle \langle \{ \xi_i \} |$.
Insertion of the identity twice into
$Z_{\bf P}$ yields
\begin{displaymath}
Z_{\bf P} = \frac{1}{N} \sum_{\bf r,r'} 
e^{i{\bf P(r'-r)}} \!\! \int_{-\infty}^{\infty} \!\!\!\!\!
D\xi D\xi' \langle {\bf r'}, \{ \xi'_i \} 
| e^{- \beta H} | \{ \xi_i \}, {\bf r} \rangle \cdot W
\end{displaymath}
\vspace{-0.5cm}
\begin{equation}
W = \sum_{\bf Q} e^{- i {\bf Q(r'-r)}} 
\sum_{n^{\bf Q}_{ph} } 
\langle \{ \xi_i \} | n^{\bf Q}_{ph} \rangle  
\langle n^{\bf Q}_{ph} | \{ \xi'_i \} \rangle .
\label{five}
\end{equation}

The quantity $W$ can now be simplified by the following
argument. Let us expand configuration $| \{ \xi_i \} \rangle$
in a Fourier series over states with
definite wave vectors ${\bf q}$:  
$| \{ \xi _i \} \rangle = N^{-1/2}
\sum_{\bf q} a_{\bf q} e^{i {\bf q R_i }} |{\bf q} \rangle$.
By $|{\bf q} \rangle$ we mean a state in which ion displacements
are arranged in a plane wave with wave vector ${\bf q}$
and the unit amplitude. 
Now comes an important observation which is
crucial for the whole method. {\em The only component
of the latter expansion which survives projection
onto $\langle n^{\bf Q}_{ph} |$ is the one with $\bf q=Q$.} 
Although $|{\bf q} \rangle$ is a wave of classical
displacements with no definite phonon occupation numbers
its expansion in quantum states $| n_{ph} \rangle$ would
contain only those with total quasi-momentum 
$\bf Q=q$ (this is simply because both the classical
and the quantum states must belong to the same 
irreducible representation of the translation group).
It is convenient to take care of this property
by introducing an extra summation over the lattice      
$\delta_{\bf Q,q} = N^{-1} \sum_{\bf m} e^{i{\bf m(Q-q)}}$.
Now configurations $\langle \{ \xi_i \} |$ and 
$| \{ \xi'_i \} \rangle$ are automatically projected
on the subspace with total quasi-momentum ${\bf Q}$ and
the sum $\sum_{n^{\bf Q}_{ph}} |n^{\bf Q}_{ph}\rangle
\langle n^{\bf Q}_{ph}|$ in Eq.(\ref{five}) may be
replaced by the unity operator.
With these transformations $W$ takes the form
\begin{displaymath}
W = \frac{1}{N^3} \sum_{\bf Q} e^{- i {\bf Q(r'-r)}}
\sum_{\bf m,m'} e^{i {\bf Q(m'-m)}} \times \makebox[2.cm]{}
\end{displaymath}
\vspace{-0.5cm}
\begin{displaymath}
\makebox[1.cm]{}
\times \sum_{\bf q,q'} a^{\ast}_{\bf q} a'_{\bf q'}
e^{-i{\bf q(R_i-m)}} e^{i{\bf q'(R_i-m')}} 
\langle {\bf q} | {\bf q'} \rangle =
\end{displaymath}
\vspace{-0.5cm}
\begin{displaymath}
=\frac{1}{N^2} \sum_{\bf m,q,q'} a^{\ast}_{\bf q} a'_{\bf q'}
e^{-i{\bf q(R_i-m)}} e^{i{\bf q'(R_i-m+r'-r)}}
\langle {\bf q} | {\bf q'} \rangle =
\end{displaymath}
\vspace{-0.5cm}
\begin{equation}
=\frac{1}{N} \! \sum_{\bf m} \langle \{ \xi_{i-{\bf m}} \} |
\{ \xi'_{i-{\bf m+r'-r}} \} \rangle  
= \prod_{i=1}^N \delta(\xi_i - \xi'_{i+{\bf r'-r}} ) .
\label{six}
\end{equation}
Deriving this result we have first summed over 
${\bf Q}$, then transformed from $|{\bf q} \rangle$
states back to $| \{ \xi_i \} \rangle$ states and
finally used normalization of the latter.  
Substitution of Eq.(\ref{six}) into Eq.(\ref{five})
and integration over $\{ \xi'_i \}$ leads now to the
final result
\begin{equation}
Z_{\bf P} = \frac{1}{N} \sum_{\bf r,r'}
e^{i{\bf P(r'-r)}} \!\! \int_{-\infty}^{\infty} \!\!\!\!\!
D\xi \, \langle {\bf r'}, \{ \xi_{i-{\bf r'+r}} \}
| e^{- \beta H} | \{ \xi_i \}, {\bf r} \rangle . 
\label{seven}
\end{equation}

Thus, unlike the total partition function,
the partial partition function with fixed ${\bf P}$
is a sum over all imaginary-time trajectories with
{\em open} boundary conditions. Moreover, the
boundary conditions for the electron and the phonons
are correlated. If the final position of the electron 
(i.e. at imaginary time $\tau=0$)
is shifted from the initial one (at $\tau=\beta$) by 
${\scriptstyle \triangle} {\bf r = r'-r}$ then 
the final phonon configuration should be obtained
from the initial one by shifting the latter by the same 
${\scriptstyle \triangle} {\bf r}$ along the lattice.   
Each $Z_{\bf P}$ receives contributions from trajectories
with all possible ${\scriptstyle \triangle} {\bf r}$.
Obviously, Eq.(\ref{seven}) satisfies the condition
$\sum_{\bf P} Z_{\bf P} = Z$.

The importance of the conservation of total quasi-momentum
in the polaron problem was first recognized by 
Lee, Low and Pines \cite{Lee} who used a canonical
transformation to eliminate electron coordinates
from the problem. Here we follow a different strategy,
namely to eliminate phonon degrees of freedom in the
spirit of the De Raedt and Lagendijk's method.
The resulting single-electron problem can be
simulated by Monte Carlo. Unfortunately,
the complex factor $e^{i {\bf P(r'-r)}}$ appearing
in Eq.(\ref{seven})
makes it impossible to simulate the system at
arbitrary ${\bf P}$ due to the complex trajectory weight.
However, {\em there is no sign-problem at} ${\bf P}=0$.   
This fact may be used, for instance, for more
accurate estimation of polaron ground state
energy from the Monte Carlo method.

In this paper we would like to point out the 
possibility of calculating polaron effective mass,
which arises from equation (\ref{seven}).
One can define the partial internal energy 
in the usual manner as 
$U_{\bf P} = - Z^{-1}_{\bf P} \partial Z_{\bf P}/\partial \beta$. 
In the limit of zero temperature $U_{\bf P}$ is reduced
to the lowest eigenvalue with quasi-momentum ${\bf P}$.
Then the inverse effective mass is (up to factor $\hbar^2$)
a second derivative of $U_{\bf P}$ with respect to
a chosen component of ${\bf P}$:  
$1/m^{\ast}_{\alpha} = \partial^2 
U_{\bf P}/\partial P^2_{\alpha} |_{{\bf P}=0}$.
From Eq.(\ref{seven}) we have
\begin{equation}
\frac{1}{m^{\ast}_{\alpha}} \propto - \left(
\left\langle (r'_{\alpha}-r_{\alpha})^2 E \right\rangle -
\left\langle (r'_{\alpha}-r_{\alpha})^2 \right\rangle 
\left\langle E \right\rangle \right)|_{{\bf P}=0} 
\label{eight}
\end{equation}
where $E$ is the value of the energy estimator
calculated on the trajectory which begins at
${\bf r}$ and ends at ${\bf r'}$ and
$\langle \ldots \rangle$ means the Monte Carlo 
average over an ensemble of such trajectories,
e.g. $\langle E \rangle|_{\bf P} = U_{\bf P}$. Thus,
{\em the polaron effective mass is inversely proportional
to the covariance of energy of the trajectory and the 
square distance between ends of the trajectory.} 
Its value can be obtained directly from Monte Carlo
simulations of the equilibrium system with open
boundary conditions at ${\bf P}=0$.

We point out that, on the semi-intuitive level, 
the relation between the width of the electron
trajectory and the polaron effective mass  
was used previously to demonstrate 
the increase of the latter \cite{Hirsch_one}.
In the present paper, however, such a relation, 
Eq.(\ref{eight}), is derived rigorously.

In order to test this method we now consider
the one-dimensional Holstein polaron problem \cite{Holstein}.
In this model the phonon subsystem is a chain of 
non-interacting oscillators with frequency $\omega$
and the electron interacts linearly only with  
the oscillator it currently occupies.
The model hamiltonian reads 
\begin{equation}
H = H_1 + H_2 + H_3
\label{nine}
\end{equation}
\vspace{-0.7cm}
\begin{displaymath}
H_1 = -t \sum_i
\left( c^{\dagger}_{i+1} c_i + c^{\dagger}_i c_{i+1} \right) ;
\qquad H_2 = \sum_i  \frac{\hat p^2_i}{2m} 
\end{displaymath} 
\vspace{-0.7cm}
\begin{displaymath}
H_3 = \sum_i \frac{m\omega^2}{2} \hat \xi^2_i -
\tilde g \sum_i c^{\dagger}_i c_i \, \hat \xi_i .
\end{displaymath}
Here $H_1$ represents electron kinetic energy,
$t$ being the nearest-neighbour hopping amplitude,
$m$ is the oscillator reduced mass, $\xi_i$ is the
displacement of $i$-th oscillator, 
$\hat p_i= -i \hbar \partial /\partial \xi_i$ and
$\tilde g$ in the electron-phonon coupling constant.
The rest of the notation is standard. 
In what follows periodic
boundary conditions in real space are assumed.

To be able to use the Monte Carlo method
one needs a path-integral representation of the
matrix element $G$ which appears in Eq.(\ref{seven}).
Following the standard procedure \cite{Hirsch_one,DeRaedt} 
we divide the
imaginary-time dimension (whose extent is $\beta$)
into $M \gg 1$ intervals,
insert the resolution of the identity $M-1$ times,
then use the Trotter decomposition and   
evaluate all the matrix elements of the operator
$e^{-(\beta/M) H}$ to get 
\begin{displaymath}
G=c_1 \! \sum_{\{ x_j \}} \int^{\infty}_{-\infty} \!
\left[ \prod^N_{i=1} \! \prod^{M-1}_{j=0} \! d\xi_{ij} \right]
e^{- S_{ph}} \left[ \prod^{M-1}_{j=0} \! I(x_{j+1}-x_j) \right]
\end{displaymath}
\vspace{-0.5cm}
\begin{equation}
I(x_{j+1}-x_j) = \frac{1}{N} \sum_{k_n}
\cos{\left[ k_n (x_{j+1}-x_j) \right]} \, e^{2\tau \cos{k_n}} .
\label{ten}
\end{equation}
Here $\{ x_j \} $ is an electron trajectory in 
imaginary time with boundary conditions
$x_0 = x, x_M = x'$; $k_n$ are single-particle momenta
allowed in a chain of $N$ sites: $k_n=2\pi n/N$, $n=0, \ldots N-1$;
$\tau \equiv \beta t/M$ and $S_{ph}$ is the phonon
action \cite{Hirsch_one,DeRaedt}  
\begin{equation}
S_{ph} \! = \! \sum^N_{i=1} \!\! \sum^{M-1}_{j=0} \!\!
\left[ \frac{m}{2\tau \hbar^2} (\xi_{i,j+1} \! - \! \xi_{ij})^2
\! + \! \tau \frac{m\omega^2}{2} \xi^2_{ij} - 
\tau \tilde g \xi_{ij} \delta_{i, x_j} \right]
\label{eleven}
\end{equation}
In Ref.\cite{DeRaedt} polaron thermodynamics were
studied. There the following result was obtained
for the integral over $\xi$s with
periodic boundary conditions $\xi_{iM}=\xi_{i0}$
\begin{equation}
\int^{\infty}_{-\infty} D\xi \ldots 
= Z_{ph} \exp \left[ \sum^{M-1}_{j=0} \sum^{M-1}_{j'=0} 
F(j-j') \, \delta_{x_j, x_{j'}} \right]
\label{twelve}
\end{equation}
where $Z_{ph}$ is the partition function of free phonons
and $F(j-j')$ is the memory function
\begin{equation}
F(j-j') = \frac{\tau^3 g^2}{4M} \sum^{M-1}_{l=0}
\frac{\cos{\frac{2\pi l}{M}(j-j')}}{1-\cos{\frac{2\pi l}{M}}+
\frac{\tau^2 \tilde \omega^2}{2}}
\label{thirteen}
\end{equation}
with $g^2 \equiv \hbar^2 \tilde g^2/m t^3$ and
$\tilde \omega \equiv \hbar \omega/t$.
In our case one has to perform the integration over
$\xi$s with twisted boundary conditions
$\xi_{i+x'-x,M}=\xi_{i,0}$. Having done this, however,
we found that, as long as the condition 
$\beta \hbar \omega \gg 1$ is satisfied, the final 
result for $x'-x \neq 0$ acquires only exponentially
small corrections relative to 
Eqs.(\ref{twelve})-(\ref{thirteen}). 
The details of the calculation are cumbersome
and will be published elsewhere \cite{Comment}.

Thus, in the low-temperature limit the path-integral
representation of the partial partition function
for the model (\ref{nine}) has the form
\begin{equation}
Z_P = \frac{1}{N} \sum_{\{ x_j \} } 
e^{i P (x'-x)} \, \rho(\{ x_j \} ) 
\label{fourteen}
\end{equation}
\vspace{-0.7cm}
\begin{displaymath}
\rho(\{ x_j \} ) \! = \! \left[ \prod^{M-1}_{j=0}
I(x_{j+1}-x_j) \right] 
e^{\sum^{M-1}_{j,j'=0} F(j-j') \delta_{x_j,x_{j'}}} . 
\end{displaymath}
Introducing the definition 
$\langle A \rangle = Z^{-1}_0 \sum_{\{ x_j \}} A \rho(\{ x_j \} )$
the expression for the effective mass can be brought
to the form
\begin{equation}
\frac{m_0}{m^{\ast}} = - \frac{1}{2}
\left( \langle (x'-x)^2 E \rangle -
\langle (x'-x)^2 \rangle \langle E \rangle \right)
\label{sixteen}
\end{equation}
\vspace{-0.5cm}
\begin{displaymath}
E = - \frac{1}{M} \!\! \sum^{M-1}_{j=0} \! \sum_{\delta = \pm 1}
\! \frac{I(x_{j+1}-x_j+\delta)}{I(x_{j+1}-x_j)}
-\frac{1}{M} \!\! \sum^{M-1}_{j,j'=0}
\frac{\partial F}{\partial \tau}
\delta_{x_j,x_{j'}}
\end{displaymath}
where $m_0 = \hbar^2/2ta^2$ is the bare electron mass
($a$ is the lattice spacing).

The inverse effective mass was calculated by the 
Metropolis Monte Carlo algorithm for a chain of 
$N=1024$ oscillators.
The frequency was $\tilde \omega = 1$
and the inverse temperature was taken 
$\beta = 15 \, t^{-1}$ to make
sure the low-temperature limit is satisfied
and Eq.(\ref{fourteen}) is valid. The number of
imaginary-time slices was varied from $M=120$ to 
$M=180$ to detect a possible $M$-dependence of
the simulation results. We {\em observed} such
a dependence in the small-polaron regime $g \geq 2.5$.
For these values of $g$ a $1/M^2$-scaling was
used to extrapolate to $M=\infty$.
In the beginning of each series of
measurements we used $50,000 \cdot M$ single-particle 
steps to warm up the system. After
that consecutive measurements were taken every 
$M$ steps. For each set of parameters 
from six to nine series of $2\cdot 10^6$ and 
$3\cdot 10^6$ measurements were made.

\begin{figure*}
\begin{center}
\leavevmode
\hbox{%
\epsfxsize=8.6cm
\epsffile{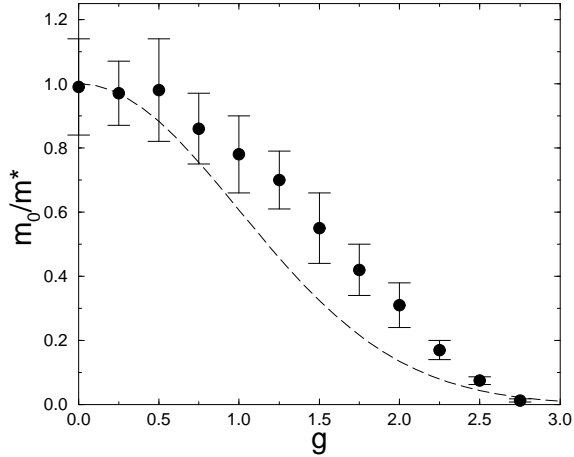}
}
\end{center}
\vspace{-0.5cm}
\caption{Inverse polaron effective mass vs 
coupling constant for the one-dimensional Holstein
model. Phonon frequency $\tilde \omega = 1$ and the inverse
temperature is $\beta = 15 \, t^{-1}$. The dashed line is 
$m_0/m^{\ast} = \exp(-g^2/2\tilde \omega^3)$.}
\end{figure*}

Results of the simulations averaged over
the different series are shown in Fig.1.
The effective mass increases exponentially
with the coupling constant, reaching 
$m^{\ast}\sim 100 \, m_0$ in the small polaron
regime $g>2.5$. In spite of the large number of measurements 
statistical errors are not small but decrease
with $g$. This reflects the nature of the object 
simulated. In the absence of electron-phonon coupling
the electron trajectory is very flexible and
its ends fluctuate strongly. As the interaction
is turned on the trajectory becomes more and more
rigid. This is because trajectories with straight
segments acquire exponentially large weights (see
Eq.(\ref{fourteen})) and dominate the  
sampling. Also shown in Fig.1 is the result of
Ref.\cite{Tyablikov} (see also 
\cite{Appel,Alexandrov_one,Holstein}) 
for the inverse effective mass (or the polaron
bandwidth) $m_0/m^{\ast} = \exp(-g^2/2\tilde \omega^3)$.
In the intermediate-coupling regime ($ 1.0 < g < 2.5$)
at $\tilde \omega = 1.0$
the Monte-Carlo polaron mass is lighter than
$\exp(g^2/2)$, which is in agreement with exact
diagonalization of finite clusters \cite{Alexandrov_three}.
 
In conclusion, we have developed a Monte Carlo algorithm
for calculating polaron effective mass. It is based
on the representation of the partition function with
fixed quasi-momentum as a path-integral with
open boundary conditions in imaginary time.   
The boundary conditions for the electron and phonon
subsystem are correlated. After analytical elimination
of phonon degrees of freedom the resulting
single-electron system with retarded self-interaction
can be simulated by Monte Carlo at zero total quasi-momentum.
The polaron effective mass turns out to be 
inversely proportional to the covariance of the energy of 
the electron trajectory and the
square distance between ends of the trajectory. 
The method does not impose any particular limitations
on the size and dimensionality of the system and
on values of the model parameters provided a
low enough temperature can be reached to satisfy the
condition $\beta \hbar \omega \gg 1$. We have 
tested the method on the one-dimensional Holstein
model and obtained physically reasonable results.
Statistical errors do not appear to be small and 
large statistics might be required to get stable numerical
results.

We are thankful to Prof. A.\,S.\,Alexandrov for attracting
our attention to this problem and to V.\,Kabanov and
E.\,Klepfish for numerous and helpful discussions.


\begin{references}
\bibitem{Comment}
The integration was done by the application of the method
of Ref.\cite{Feynman} to the discrete-time case. 
The first step is to find a classical trajectory
which minimizes the phonon action.
\end{references}
\end{document}